\documentclass[12pt]{article}
\usepackage{amssymb}
\usepackage[dvips]{epsfig}

\newcommand{\be}{\begin{equation}}
\newcommand{\ee}{\end{equation}}
\def\bea{\begin{eqnarray}}
\def\eea{\end{eqnarray}}

\newcommand{\bn}{\begin{eqnarray}}
\newcommand{\en}{\end{eqnarray}}

\newcommand{\oc}{{\overline{c}}}

\newcommand{\p}{\partial}

\newcommand{\nn}{\nonumber}

\newcommand{\tis}{\tilde{s}}
\newcommand{\tir}{\tilde{r}}
\newcommand{\no}{\noindent}

\newcommand{\cL}{{\cal{L}}_}

\newcommand{\s}{\,\,\,\,}
\def\bea{\begin{eqnarray}}
\def\eea{\end{eqnarray}}

\newcommand{\beq}{\begin{eqnarray}}
\newcommand{\eeq}{\end{eqnarray}}
\begin{document}

\title{\textbf{ Unitarity of
spin-2 theories with linearized Weyl symmetry in $D=2+1$}}
\author{D. Dalmazi   \\
\textit{{UNESP - Campus de Guaratinguet\'a - DFQ} }\\
\textit{{Av. Dr. Ariberto Pereira da Cunha, 333} }\\
\textit{{CEP 12516-410 - Guaratinguet\'a - SP - Brazil.} }\\
\textsf{E-mail: dalmazi@feg.unesp.br }}
\date{\today}
\maketitle

\begin{abstract}

Here we prove unitarity of the recently found fourth-order
self-dual model of spin-2 by investigating the analytic structure
of its propagator. The model describes massive particles of
helicity $+2$ (or -$2$) in $D=2+1$ and corresponds to the
quadratic truncation of a higher derivative topologically massive
gravity about a flat background. It is an intriguing example of a
theory where a term in the propagator of the form $1/\lbrack\Box^2
(\Box - m^2)\rbrack $  does not lead to ghosts. The crucial role
of the linearized Weyl symmetry in getting rid of the ghosts is
pointed out. We use a peculiar pair of gauge conditions which fix
the linearized reparametrizations and linearized Weyl symmetries
separetely.

\end{abstract}

\newpage

\section{Introduction}

It is commonly believed that higher derivative theories, though
improve the ultraviolet behavior of field theories, lead to
violations of unitarity. In the case of spin-2 particles in
$D=2+1$ an interesting exception is the third order topologically
massive gravity (TMG) of \cite{djt} which describes a massive
particle of helicity $+2$ (or $-2$). A quantum and covariant way
of understanding the absence of instabilities, about a flat
background, in this theory is to relate its linearized version,
via master action approach \cite{dj}, to the first order
(ghost-free) self-dual model ($S_{SD}^{(1)}$) of \cite{aragone} by
the addition of trivial (no particle content) mixing terms. The
procedure is such that the original physical content of
$S_{SD}^{(1)}$ is preserved in the dual theory and no extra
propagating poles show up. The complete master action is given in
\cite{prd2009}.

Another interesting exception is the new massive gravity theory of \cite{bht} (BHT theory) which contains a
second-order Einstein-Hilbert action with ``wrong'' sign and a fourth-order term with  curvature squares with
fine tuned coefficients (K-term). This theory describes a parity doublet of massive particles of helicities $+2$
and $-2$ in $D=2+1$. Also in this case, there is a master action \cite{bht} relating this model to the
ghost-free Fierz-Pauli theory by adding trivial mixing terms (linearized Einstein-Hilbert action) such that no
ghosts are expected. Indeed, the unitarity of the BHT theory has been shown in \cite{oda} by an explicit
analysis of the analytic structure of propagator. In this work we are concerned with a new fourth-order
self-dual model ($S_{SD}^{(4)}$) of spin-2 deduced in \cite{sd4} by a Noether gauge embedment procedure and
suggested also in \cite{andringa}. The new model has been shown to be dual at quantum level to $S_{SD}^{(1)}$.
The explicit dual map is given in \cite{sd4} where a master action relating it to the linearized topologically
massive gravity of \cite{djt} is presented. Collecting the results of \cite{sd4} and \cite{prd2009} one can show
that $S_{SD}^{(4)}$ stems from $S_{SD}^{(1)}$ by the addition of trivial (non-propagating) mixing terms. Based
on this explicitly covariant argument we expect only one massive particle in the spectrum of $S_{SD}^{(4)}$. The
aim of this work is to present a detailed calculation of the residues about the poles of the propagator of the
$S_{SD}^{(4)}$ theory, thus confirming our expectations. The crucial role of the linearized Weyl symmetry is
made clear. In the next section we start with a bit more general Lagrangian and study also the subcases
corresponding to the pure K-term, addressed in \cite{deserpre} from another point of view, and the pure
gravitational Chern-Simons term of \cite{djt}, both at linearized level about a flat background.

\section{Covariant gauge fixing}

We start with the action of the self-dual model of
\cite{sd4,andringa} which corresponds to the linearized version of
a higher derivative topologically massive gravity:

\be S (a,b) = \int d^3 \, x \left\lbrack b \sqrt{-g}\left( R_{\mu\nu} R^{\nu\mu} - \frac 38 R^2 \right) - \frac
a2 \epsilon^{\mu\nu\rho}\Gamma_{\mu\gamma}^{\epsilon}\left( \p_{\nu}\Gamma_{\epsilon\rho}^{\gamma} + \frac 23
\Gamma_{\nu\delta}^{\gamma}\Gamma_{\rho\epsilon}^{\delta} \right)\right\rbrack_{hh} \quad .\label{sab} \ee

\no Where $(a,b)$ are arbitrary real constants with mass dimension
-1 and -2 respectively. More explicitly, we can write the
corresponding Lagrangian density:

\be \cL{ } (a,b) = \frac b4 h_{\lambda\mu} \Box^2 \left(
\theta^{\lambda}_{\s\alpha} \theta^{\mu}_{\s\beta} -
\frac{\theta^{\lambda\mu}\theta_{\alpha\beta}}{2}\right)
h^{\alpha\beta} \, + \, \frac a2 h_{\lambda\mu}
E^{\lambda}_{\s\alpha}\Box \, \theta^{\mu}_{\s\beta} \,
h^{\alpha\beta} \quad .  \label{lab} \ee

\no We use the following definitions throughout this work:

\bea g_{\alpha\beta} &=& \eta_{\alpha\beta} + h_{\alpha\beta} \quad ,
\label{hab} \\
\Box \theta_{\alpha\beta} &=& \Box \left( \eta_{\alpha\beta} - \omega_{\alpha\beta} \right) \quad , \quad
\omega_{\alpha\beta} = \frac{\p_{\alpha}\p_{\beta}}{\Box} \quad , \label{t} \\
E_{\alpha\beta} &=& \epsilon_{\alpha\beta\gamma}\p^{\gamma} \quad
, \quad  \hat{E}_{\alpha\beta} =
\frac{E_{\alpha\beta}}{\sqrt{\Box}} \label{e} \eea

\no  with $\eta_{\alpha\beta}=(-,+,+) \, $ and $ \, \epsilon_{012}=1$. All second rank tensors are symmetric,
e.g., $h_{\alpha\beta} = h_{\beta\alpha}$ except $E_{\alpha\beta} = - E_{\beta\alpha}$ (and
$\hat{E}_{\alpha\beta}$). The action $S(a,b)$ is invariant under linearized reparametrizations and linearized
Weyl symmetries respectively:

\bea \delta_{\Lambda} h_{\mu\nu} &=& \p_{\mu}\Lambda_{\nu} +
\p_{\nu} \Lambda_{\mu} \quad ,
\label{dl}\\
\delta_{\phi} h_{\mu\nu} &=& \phi \, \eta_{\mu\nu} \quad ,
\label{df} \eea

\no In order to obtain the propagator from (\ref{lab}) one first try to add a gauge fixing term of the de
Donder-like form: $\cL{GF1} = f^{\mu} f_{\mu}/2\lambda_1$ associated with the gauge condition:

\be f_{\mu} = r \, \p^{\nu} h_{\mu\nu} + s \, \p_{\mu} h = 0 \quad
. \label{gc1} \ee

\no With $(r,s)$ real constants and $h=h_{\mu}^{\s\mu}$.  However,
if we choose $\Lambda_{\mu} = A \, \p_{\mu}\Omega$ and $\phi = B
\, \Box \,\Omega$ in (\ref{dl}) and (\ref{df}) we find the gauge
transformation:

\be \delta_G f_{\mu} \equiv \left(\delta_{\phi} +
\delta_{\Lambda}\right) f_{\mu} = \left\lbrack 2 A \left( r + s
\right) + B\left( r + 3\, s \right) \right\rbrack \Box \, \p_{\mu}
\Omega \quad . \label{g} \ee

\no Since $\Omega$ is an arbitrary function, it is clear that for
any choice of $r$ and $s$ we can always find a real pair
$\left(A,B\right)$ such that $2 A \left( r + s \right) + B\left( r
+ 3\, s \right)=0$. Thus, the gauge condition (\ref{gc1}) leaves a
residual symmetry. Consequently, another (scalar) gauge fixing
condition $f=0$ and a further gauge fixing term $\cL{GF2} =
f^2/2\lambda_2$ are absolutely necessary. Since $\cL{ } (a,b)$ is
a pure higher derivative theory we have found natural to use the
higher-order gauge condition:

\be f = \tilde{r} \, \p^{\mu}\p^{\nu} h_{\mu\nu} + \tilde{s}
\,\Box \, h = 0 \quad . \label{gc2} \ee

\no The real constants $(\tir,\tis)$ are arbitrary except for the
forbidden choice $(\tir,\tis)=(r,s)$ which makes the second gauge
condition (\ref{gc2}) not independent of the first one
(\ref{gc1}). Although for the linearized theory considered here
the ghosts decouple from the physical field $h_{\mu\nu}$, it may
be useful in a more general situation to consider the following
argument based on the Faddeev-Popov Lagrangian ($\cL{FP}$) in
order to find a convenient choice for the couple $(\tir,\tis)$.
Namely, from the gauge transformations we have:

\be \cL{FP} = \oc^{\mu} \left(\delta_{\phi=c} + \delta_{\Lambda_{\alpha}=c_{\alpha}}\right)f_{\mu} \, + \, \oc
\left(\delta_{\phi=c} + \delta_{\Lambda_{\alpha}=c_{\alpha}}\right)f \quad . \label{lfp1} \ee

\no From (\ref{gc1}),(\ref{g}) and (\ref{gc2}) it is clear that
only for $s=-r/3$ and $\tis = - \tir$ the Weyl ghosts and
anti-ghosts decouple from the reparametrization ghosts and
anti-ghosts. This means that the gauge conditions (\ref{gc1}) and
(\ref{gc2}) fix separately the linearized reparametrization and
linearized Weyl symmetries without any mixing. Moreover, the
Faddeev-Popov Lagrangian becomes purely second-order: $\cL{FP} =
\oc^{\mu}\left(\Box \eta_{\mu\nu} +
\p_{\mu}\p_{\nu}/3\right)c^{\nu} + 2 \oc \, \Box \, c$ with the
global symmetry $\delta \oc^{\nu} = l \, \p^{\nu} \oc \; \delta \,
c = 2 \, l\, \p^{\nu}c_{\nu}/3 $ where $l$ is a constant. The
above gauge  choices lead to the following gauge fixing term which
will be used henceforth:

\be \cL{GF} = \cL{GF1} + \cL{GF2} = \frac 1{2\lambda_1}(
\p^{\mu}h_{\mu\nu} - \frac 13 \p_{\nu} h )^2 + \frac
1{2\lambda_2}\left( \Box \, h - \p^{\mu}\p^{\nu} h_{\mu\nu}
\right)^2 \quad . \label{gf} \ee

\no We can rewrite $\cL{ }(a,b) + \cL{GF}$ in terms of spin projection operators:

\be  \cL{ }(a,b) + \cL{GF} = h_{\lambda\mu}{\cal
O}^{\lambda\mu}_{\s\s\alpha\beta} h^{\alpha\beta} \quad .
\label{total} \ee

\no Where:

\bea {\cal O}^{\lambda\mu}_{\s\s\alpha\beta} = &{ }& \Box \left\lbrack
 \left(\frac b4 \Box +  \frac a2 \sqrt{\Box} \right)
P_{+SS}^{(2)} + \left(\frac b4 \Box - \frac a2 \sqrt{\Box} \right) P_{-SS}^{(2)} + \frac{\lambda_1 \Box -
\lambda_2}{4\lambda_1\lambda_2} P_{SS}^{(1)} \right.  \nn\\
 &-& \left. \frac 2{9\lambda_1} P_{WW}^{(0)} +
\frac{\sqrt{2}}{9\lambda_1}\left( P_{SW}^{(0)} +
P_{WS}^{(0)}\right) + \left( \frac{\Box}{\lambda_2} - \frac 1{9\,
\lambda_2} \right) P_{SS}^{(0)}
\right\rbrack^{\lambda\mu}_{\s\s\alpha\beta} \quad . \label{ocal}
\eea

\no Where, following the notation of appendix B of
\cite{ariasphd}, slightly modified, the projection operators of
spin 2, 1 and 0 introduced above are given respectively by:

\be \left( P_{\pm SS}^{(2)} \right)^{\lambda\mu}_{\s\s\alpha\beta}
= \frac 14 \left(
\theta_{\pm\s\alpha}^{\s\lambda}\theta^{\mu}_{\s\beta} +
\theta_{\pm\s\alpha}^{\s\mu}\theta^{\lambda}_{\s\beta} +
\theta_{\pm\s\beta}^{\s\lambda}\theta^{\mu}_{\s\alpha} +
\theta_{\pm\s\beta}^{\s\mu}\theta^{\lambda}_{\s\alpha} -
\theta^{\lambda\mu} \theta_{\alpha\beta}
 \right) \quad , \label{ps2} \ee

\be \left( P_{SS}^{(1)} \right)^{\lambda\mu}_{\s\s\alpha\beta} =
\frac 12 \left(
\theta_{\s\alpha}^{\lambda}\,\omega^{\mu}_{\s\beta} +
\theta_{\s\alpha}^{\mu}\,\omega^{\lambda}_{\s\beta} +
\theta_{\s\beta}^{\lambda}\,\omega^{\mu}_{\s\alpha} +
\theta_{\s\beta}^{\mu}\,\omega^{\lambda}_{\s\alpha}
 \right) \quad , \label{ps1} \ee

\be \left( P_{SS}^{(0)} \right)^{\lambda\mu}_{\s\s\alpha\beta} =
\frac 12 \, \theta^{\lambda\mu}\theta_{\alpha\beta} \quad , \quad
\left( P_{WW}^{(0)} \right)^{\lambda\mu}_{\s\s\alpha\beta} =
\omega^{\lambda\mu}\omega_{\alpha\beta} \quad , \label{psspww} \ee

\be \left( P_{SW}^{(0)} \right)^{\lambda\mu}_{\s\s\alpha\beta} =
\frac 1{\sqrt{2}}\,  \theta^{\lambda\mu}\omega_{\alpha\beta} \quad
, \quad  \left( P_{WS}^{(0)}
\right)^{\lambda\mu}_{\s\s\alpha\beta} = \frac 1{\sqrt{2}}\,
\omega^{\lambda\mu}\theta_{\alpha\beta} \quad , \label{pswpws} \ee

\no Where

 \be \theta_{\pm\s\beta}^{\s\alpha} = \frac 12 \left( \theta^{\alpha}_{\s\beta} \pm \hat{E}^{\alpha}_{\s\beta}
 \right) \quad . \label{tpm}\ee

\no Note that the factor $\sqrt{\Box}$ cancel out in (\ref{ocal}).
Taking\footnote{We use the notation $\left(P \cdot
Q\right)^{\lambda\mu}_{\s\s\alpha\beta}=
P^{\lambda\mu}_{\s\s\gamma\delta}
Q^{\gamma\delta}_{\s\s\alpha\beta}$.} into account $\,P_{+
SS}^{(2)} \cdot P_{-SS}^{(2)}=0$, $P_{\pm SS}^{(2)} \cdot
P_{SS}^{(1)}=0\,$, $\, P_{\pm SS}^{(2)} \cdot P_{SS}^{(0)}=0 \,$ ,
$P_{ SS}^{(1)} \cdot P_{SS}^{(0)} =0$, the spin-0 algebra:
$P_{IJ}^{(0)}\cdot P_{KL}^{(0)} = \delta_{JK} P_{IL}^{(0)}$  and
the representation of the symmetric identity

\be \left( 1_S \right)^{\lambda\mu}_{\s\s\alpha\beta} = \frac 12
\left(\delta^{\lambda}_{\s\alpha} \delta^{\mu}_{\s\beta} +
\delta^{\mu}_{\s\alpha} \delta^{\lambda}_{\s\beta} \right) =
\left\lbrack P_{+SS}^{(2)}+ P_{-SS}^{(2)} + P_{SS}^{(1)} +
P_{SS}^{(0)} + P_{WW}^{(0)}
\right\rbrack^{\lambda\mu}_{\s\s\alpha\beta} \quad ,
\label{identity} \ee

\no one can solve the equation $ {\cal O} \cdot {\cal O}^{-1} =
1_S $ and find:

\bea \left({\cal O}^{-1}\right)^{\lambda\mu}_{\s\s\alpha\beta} = &{ }& \left\lbrace \frac 4b \left\lbrack
\frac{P_{+SS}^{(2)}}{\Box \left(\Box + m \sqrt{\Box} \right)} + \frac{P_{-SS}^{(2)}}{\Box \left(\Box - m
\sqrt{\Box} \right)} \right\rbrack + \frac{4\lambda_1\lambda_2}{\Box \left(\lambda_1 \Box -
\lambda_2\right)}P_{SS}^{(1)} + \frac{\lambda_2}{\Box^2} P_{SS}^{(0)} \right. \nn\\
&+& \left. \left(\frac{\lambda_2}{\Box} - 9 \lambda_1 \right) \frac{P_{WW}^{(0)}}{2\, \Box} +
\frac{\lambda_2}{\sqrt{2} \, \Box^2} \left( P_{SW}^{(0)} + P_{WS}^{(0)}\right)
\right\rbrace^{\lambda\mu}_{\s\s\alpha\beta} \label{ocalm} \eea

\no where $m= 2\, a/b$.

 Now we are ready to obtain the propagator in momentum space
saturated with transverse, symmetric and traceless (Weyl symmetry)
sources along the lines of \cite{nieu}. Using the reality
condition of the sources in coordinate space, the desired result
in the momentum space, called henceforth $A(k)$, can be written
as:

\be A(k) = T^{*}_{\alpha\beta}(k)\left\langle \tilde{h}^{\alpha\beta}(-k)\, \tilde{h}_{\lambda\mu} (k)
\right\rangle T^{\lambda\mu}(k) \quad , \label{A} \ee

\no where $\tilde{h}_{\lambda\mu}(k)$ stand for the Fourier transform of $h_{\lambda\mu}(x)$ and the sources
must satisfy:

\bea k_{\mu} T^{\mu\nu}&=& 0  \, = \, T^{\mu\nu} k_{\nu} \label{trans} \\
T^{\mu\nu} &=& T^{\nu\mu} \label{symm} \\
T^{\mu}_{\s\mu} &=& - T^{00} + T^{11} + T^{22} = 0 \label{trace}
\eea

\no When we sandwich the operator $\left({\cal O}^{-1}\right)^{\lambda\mu}_{\s\s\alpha\beta}$ with sources
satisfying (\ref{trans}),(\ref{symm}) and (\ref{trace}), only the contributions of the spin-2 operators survive
which is of course expected since the lower spin terms are gauge dependent and should not interfere in the
analysis of the particle content of the model. Of course, we need to be careful in the neighborhood of the poles
as we will stress later. At this point it is instructive to split the spin-2 operators in even and odd parity
sectors and write:

\be \left\lbrack \left({\cal
O}^{-1}\right)^{\lambda\mu}_{\s\s\alpha\beta}\right\rbrack_{s=2} =
\frac 4b \left\lbrack \frac{P_{SS}^{(2)}}{\Box \left(\Box - m^2
\right)} - \frac{m}{\Box^{3/2} \left(\Box - m^2
 \right)}\left( P_{+SS}^{(2)}- P_{-SS}^{(2)}\right) \right\rbrack^{\lambda\mu}_{\s\s\alpha\beta}
  \quad . \label{ocalms2} \ee

\no Where the parity even spin-2 operator is defined as
$P_{SS}^{(2)}=P_{+SS}^{(2)}+ P_{-SS}^{(2)}$. Using (\ref{trans})
we derive the following identities (in momentum space
$\hat{E}_{\alpha\beta} = i\, \epsilon_{\alpha\beta\gamma}\,
k^{\gamma}/\sqrt{-k^2}$):

\bea T^{*}_{\alpha\beta}\left( P_{SS}^{(2)} \right)^{\alpha\beta}_{\s\s\lambda\mu}T^{\lambda\mu} &=&
T_{\alpha\beta}^{*} T^{\alpha\beta} \quad , \label{tteven} \\
T^{*}_{\alpha\beta}\left( P_{+SS}^{(2)} - P_{-SS}^{(2)} \right)^{\alpha\beta}_{\s\s\lambda\mu}T^{\lambda\mu} &=&
T_{\alpha\beta}^{*}\hat{E}^{\alpha}_{\s\lambda} \theta^{\beta}_{\s\mu} T^{\lambda\mu} =
T_{\alpha\beta}^{*}\hat{E}^{\alpha}_{\s\lambda} T^{\lambda\beta} \quad , \label{ttodd} \eea

\no Then, after trivial rearrangements we can write:

\be A(k) = \frac {2 \, i}{b \, m^2} \left\lbrack \left(
T_{\alpha\beta}^{*} T^{\alpha\beta} -
\frac{T_{\alpha\beta}^{*}E^{\alpha}_{\s\lambda}
T^{\lambda\beta}}{m} \right) \left( \frac 1{k^2 + m^2} -  \frac
1{k^2} \right) - \frac m{(k^2)^2} \left(
T_{\alpha\beta}^{*}E^{\alpha}_{\s\lambda} T^{\lambda\beta} \right)
\right\rbrack \quad . \label{A2} \ee

 Next, we need to calculate the imaginary part of the residue about the poles of $A(k)$. We
first analyze the massive pole. We choose the convenient frame
$k_{\mu} = (m,0,0)$, so from the transverse condition
(\ref{trans}) we have $T^{0 \mu} = 0 = T^{\mu 0} \, , \, \mu=0,1,2
$. Therefore,

\bea T_{\alpha\beta}^{*} T^{\alpha\beta} &=& \vert T_{11} \vert^2 +\vert T_{22} \vert^2 + 2 \, \vert T_{12}
\vert^2 \quad , \label{ttm} \\
T_{\alpha\beta}^{*}E^{\alpha}_{\s\lambda} T^{\lambda\beta} &=&  m
\, i \left\lbrack T_{12}^{*}\left(T_{11}-T_{22}\right) -
\left(T_{11}^{*}-T_{22}^{*}\right)T_{12} \right\rbrack \quad ,
\label{tetm} \eea

\no From (\ref{A2}),(\ref{ttm}) and (\ref{tetm}) we have:

\be {\rm Im} \, {\rm Residue} \left\lbrack
A(k)\right\rbrack_{k^2=-m^2} = \lim_{k^2\to -m^2} \left( k^2 + m^2
\right) A (k) = \frac 2{b\, m^2} \left( \vert T_{11} + i\, T_{12}
\vert^2 + \vert T_{22} - i \, T_{12} \vert^2 \right)_{k^2=-m^2}
\quad . \label{mfinal}\ee

\no Therefore we conclude that ${\rm Im} \, {\rm Residue}
\left\lbrack A(k)\right\rbrack_{k^2=-m^2} > 0$ whenever $b > 0$
which proves, in agreement with the classical canonical analysis
of \cite{andringa}, that we have, assuming of course $a \ne 0$,
one physical massive particle in the spectrum of $\cL{ } (a,b)$.
We remark that it is not necessary to use the traceless condition
on the sources to derive (\ref{mfinal}).

Next we turn to the more subtle case of the massless poles. We use
the frame $k_{\mu}=(-k_0,\epsilon,-k_0)$, which implies
$k^2=\epsilon^2$, and take afterwards the limit $\epsilon \to 0$
assuming $\epsilon/k_0 < 1 $. In fact, we need to calculate $ {\rm
Im} \, {\rm Residue} \left\lbrack A(k)\right\rbrack_{k^2=0} =
\lim_{\epsilon \to 0} \epsilon^2 \, A(k) $ in the above frame
where the transverse condition can be written as the following 3
equations:

\be k_0 \left( T^{0\mu} + T^{2\mu} \right) = \epsilon \, T^{1\mu} \, , \, \mu=0,1,2 \quad. \label{trans0} \ee

\no It is clear from (\ref{A2}) that we need to evaluate the two
quantities:  $V(k) \equiv
T_{\alpha\beta}^{*}E^{\alpha}_{\s\lambda} T^{\lambda\beta}$ and
$U(k) \equiv T_{\alpha\beta}^{*} T^{\alpha\beta}$ in the limit
$\epsilon \to 0$. We first look at $V(k)$. If we use the symmetric
condition (\ref{symm}), the transverse condition and its complex
conjugated we can write:

\bea V(k) &=& i \, \left\lbrack k_0 \, T_{\mu}^{*\s 1}\left(
T^{2\mu} + T^{0\mu} \right) - k_0 \left(T_{\mu}^{*\s 2} +
T_{\mu}^{*\s 0}\right) T^{1\mu}  + \epsilon\, T_{\mu}^{*\s 2}
T^{0\mu} -
\epsilon \, T_{\mu}^{*\s 0} T^{2\mu} \right\rbrack \quad , \label{v1} \\
&=& i \, \epsilon \left( T_{\mu}^{*\s 2} T^{0\mu} - T_{\mu}^{*\s
0} T^{2\mu} \right) \label{v2} \eea

\no So it is already clear, without  the traceless  condition
(\ref{trace}), that the term $V(k)$ will not contribute to any
residue at the simple massless pole in (\ref{A2}) but (\ref{v2})
is not enough to get rid of the massless double pole. Using
(\ref{trans0}) and the traceless condition (\ref{trace}) we can
eliminate 4 out of 6 components of the symmetric tensor
$T_{\mu\nu}$. It is convenient to choose $T^{02}$ and $T^{12}$ as
the  independent variables\footnote{Other choices may require
specific properties of some of the components of $T^{\mu\nu}$ at
$\epsilon \to 0$ in order to guarantee that all $T_{\mu\nu}$
behave smoothly at $\epsilon \to 0$. However, by using such
properties the leading behavior in (\ref{e4}) will not change}.
Explicitly, without any approximation we have:

\bea
 T^{00} &=& - \frac{(\epsilon/k_0)(1+ \epsilon^2/k_0^2)}{(1- \epsilon^2/k_0^2)} T^{12} - T^{02} \quad ,
 \label{t00}\\
 T^{01} &=& - \frac{(1+ \epsilon^2/k_0^2)}{(1- \epsilon^2/k_0^2)} T^{12} \quad , \label{t01} \\
T^{11} &=&  - \frac{\left(2\, \epsilon/k_0 \right)}{(1- \epsilon^2/k_0^2)} T^{12} \quad , \label{t11} \\
T^{22} &=& \frac{\epsilon}{k_0} T^{12} - T^{02} \quad ,
\label{t22} \eea

\no It turns out that plugging all the above formula in (\ref{v2}) we end up with the leading behavior

\be V(k)= T_{\alpha\beta}^{*}E^{\alpha}_{\s\lambda}
T^{\lambda\beta} = \frac{ i \, 2 \epsilon
\left(\epsilon^3/k_0^3\right)}{(1-\epsilon^2/k_0^2)}\left(T_{02}T_{12}^{*}
- T_{02}^* T_{12} \right) \approx C (k_0) \, \epsilon^4 + \cdots
\qquad . \label{e4} \ee

\no Where $C(k_0)$ is real but has {\it a priori} no definite
sign. Anyway, it is now clear that the massless double pole of
(\ref{A2}) does not contribute to $\lim_{\epsilon\to
0}\epsilon^2\, A(k)$ and drops out of the saturated propagator.
The linearized Weyl symmetry has played a crucial role.

As a final step we have to evaluate the quantitiy
$U(k)=T_{\alpha\beta}^{*} T^{\alpha\beta}$ at $\epsilon \to 0$. By
using (\ref{t00}),(\ref{t01}),(\ref{t11}), and (\ref{t22}) we
obtain:

\be T_{\alpha\beta}^{*} T^{\alpha\beta} = -
\frac{2(\epsilon/k_0)^2(1+ \epsilon^2/k_0^2)}{(1-
\epsilon^2/k_0^2)} \vert T_{12} \vert^2 - \frac{2\,
(\epsilon/k_0)^3}{(1- \epsilon^2/k_0^2)}\left(T_{02}T_{12}^{*} +
T_{02}^* T_{12}\right) \approx - 2
\left(\frac{\epsilon}{k_0}\right)^2 \vert T_{12} \vert^2 \quad .
\label{e2} \ee

\no So finally we get rid completely of the massless simple pole
too. We conclude that the $S_{SD}^{(4)}$ model is free of ghosts,
for $b > 0$,  and only contains one (physical) massive particle,
 of helicity $+2$ (or -$2$), depending upon
the sign of the constant $a$ which is not fixed by unitarity.

Now we comment on two interesting subcases corresponding to $a=0$
(linearized pure K-term) and $b=0$ (linearized pure gravitational
Chern-Simons term). If we take $a \to 0$ ($m = 2a/b \to 0$) in
(\ref{A2}) we obtain a massless double pole: $ A(k) = -(2i/b)
T_{\alpha\beta}^{*} T^{\alpha\beta}/(k^2)^2 $. However, due to
(\ref{e2}) we have a finite residue:

\be {\rm Im} \, {\rm Residue} \left\lbrack
A(k)\right\rbrack_{k^2=0} = \lim_{\epsilon \to 0} \epsilon^2 \,
A(k) = \epsilon^2 \left(\frac 2b \right) \frac {2
\epsilon^2}{k_0^2} \frac{\vert T_{12} \vert^2}{\epsilon^4} = \frac
{4}{b\, k_0^2}\vert T_{12} \vert^2 \quad , \label{a0} \ee

\no Therefore for a positive coefficient ($ b > 0 $) the pure
K-term is ghost-free and contains only one massless physical
particle in the spectrum, in agreement with the classical
canonical analysis of \cite{deserpre,bht2}. Regarding the second
special case of the pure $CS_3$ theory, if we take $b\to 0$
($m\to\infty$) only the last term of (\ref{A2}) survives: $A(k) =
(i/a)V(k)/(k^2)^2$. Due to (\ref{e4}) the residue vanishes:

\be \lim_{\epsilon \to 0} \epsilon^2 \, A(k) = \epsilon^2
\left(\frac ib\right) \frac{C(k_0)\epsilon^4}{\epsilon^4} = 0
\quad . \label{b0} \ee

\no which confirms the trivial (non-propagating) nature of the
pure gravitational Chern-Simons term ($CS_3$), see canonical
analysis in \cite{djtprl}.

Finally, a remark is in order regarding the covariance of our calculations. Since we have used specific
reference frames in the analysis of both massive and the massless poles we have lost explicit covariance.
However, based on similar calculations in the spin-1 Maxwell-Chern-Simons (MCS) theory of \cite{djt} we believe
that explicit covariance can be recovered in principle. In the MCS theory we saturate the propagator with
conserved currents $k_{\mu}J^{\mu} =0$ such that the amplitude contains a parity-odd term with a massless pole:
$A(k)= (i/m) J_{\mu}^* E^{\mu}_{\s\nu} J^{\nu}/k^2 + \cdots $. Where the dots stand for analytic terms at
$k^2=0$. Although, the simplest way to prove that the residue at $k^2=0$ vanishes is to choose  a convenient
frame, as we have done here, one can alternatively use the covariant identity $E^{\mu}_{\s\nu}k_{\beta} =
E^{\mu}_{\s\beta} k_{\nu} - E_{\nu\beta} k^{\mu} + k^2 \epsilon^{\mu}_{\s\nu\beta}$ from which we can easily
prove, using current conservation, that $\left( J_{\mu}^* E^{\mu}_{\nu} J^{\nu}\right)k_{\beta}/k^2 =
\epsilon^{\mu}_{\s\nu\beta}J_{\mu}^*J^{\nu} $. Since, at least, one component of $k_{\beta}$ must be
non-vanishing, the residue of $A(k)$ at $k^2=0$ is shown to vanish in a explicit covariant way. We believe that
similar identities for rank-two tensors can be used in order to make our calculations explicitly covariant
without ever using polarization vectors as in \cite{nieu}.

\section{Conclusion}

We have demonstrated here how the local linearized Weyl symmetry
can help us in getting rid of massless ghosts and double poles in
a purely higher derivative theory. The model in question
corresponds to the newly found \cite{sd4,andringa} self-dual model
which describes massive particles of helicity $+2$ (or $-2$) in
$D=2+1$. Our results agree with the classical canonical analysis
of \cite{andringa}. This case should be contrasted with the BHT
theory \cite{bht} which is also of fourth order but includes the
second order Einstein-Hilbert action with a ``wrong'' sign and
describes a massive parity-doublet of helicities $+2$ and $-2$. In
that case there are no double poles and the  residue at the
ghost-like massless pole vanishes in a quite different way
\cite{oda}. In both cases it must be mentioned that unitarity was
already expected from the master action point of view. The key
ingredient is that both Lagrangians contain a trivial term with no
particle content which allows a spectrum preserving relationship
with a lower-order model as shown in \cite{sd4} and \cite{bht}
respectively.

Here we have also shown that the case of the linearized pure
K-term contains a massless particle with non-vanishing residue in
the spectrum in agreement with the classical canonical analysis of
\cite{deserpre}, see also \cite{bht2}, despite the double pole in
the propagator.

Finally, we remark that a couple of gauge conditions (\ref{gc1})
and (\ref{gc2}) was necessary in order to obtain the propagator.
It is not clear for us how both gauge conditions could be
interpreted as linearizations of gauge conditions valid for the
full higher derivative topologically massive (HDTMG) theory. To
the best we know, the only local symmetries of HDTMG is general
coordinate invariance for which one vector gauge condition should
be expected. The same problem occurs in the pure K-theory, $a=0$
in (\ref{sab}), since the Weyl symmetry only appears after
linearization. Only in the trivial case of the pure gravitational
Chern-Simons term, $b=0$ in (\ref{sab}), the Weyl symmetry is
present in the full non-linear theory, so there is no
interpretation problem for our couple of gauge conditions.

\section{Acknowledgements}

This work is partially supported by \textbf{CNPq}. The author thanks Elias L. Mendon\c ca for several
discussions.


\begin{thebibliography}{99}


\bibitem{djt} S. Deser, R. Jackiw and S. Templeton,
Annals Phys.140:372-411,1982, Erratum-ibid.185:406,1988, Annals Phys.281:409-449,2000.

\bibitem{dj} S.Deser and R. Jackiw, Phys.Lett.B {\bf 139} (1984) 371.

\bibitem{aragone} C. Aragone and A. Khoudeir, Phys. Lett.
B{\bf173} 141 (1986).

\bibitem{prd2009} D.Dalmazi and E.L.Mendon\c ca, Phys.Rev.D\textbf{79} 045025 (2009).

\bibitem{bht} E. Bergshoeff, O. Hohm and P.K. Townsend, Phys.Rev.Lett.102:201301,2009.

\bibitem{oda} M. Nakazone and I. Oda, Prog. Theor. Phys. 121 (2009), 1389, see also arXiv:0902.3531.

\bibitem{sd4} D.Dalmazi and E.L.Mendon\c ca, `` A new spin-2 self-dual model in $D=2+1$'', arXiv:0907.5009.

\bibitem{andringa}  R. Andringa, Eric A. Bergshoeff, M. de Roo, O. Hohm, E. Sezgin and P. K. Townsend,
`` Massive 3D Supergravity'' arXiv:0907.4658.

\bibitem{deserpre} S. Deser, ``Ghost-free, finite, fourth order D=3 (alas)
gravity'', arXiv:0904.4473.

\bibitem{ariasphd} P. J. Arias, ``Spin-2 in (2+1)-dimensions'',  PhD thesis (Simon Bolivar Univ.), in
Spanish, gr-qc/9803083

\bibitem{nieu} P. van Nieuwenhuizen, Nucl. Phys. {\bf B60} 478 (1973).

\bibitem{djtprl} S. Deser, R. Jackiw and S. Templeton, Phys.Rev.Lett.48:975-978,1982.

\bibitem{bht2} E. Bergshoeff, O. Hohm and P.K. Townsend, Phys.Rev.D79:124042,2009.

\end{thebibliography}
\end{document}